\documentclass[twocolumn,preprintnumbers,amsmath,amssymb,superscriptaddress]{revtex4}

\usepackage{graphicx}
\usepackage{dcolumn}
\usepackage{bm}
\usepackage{soul}
\usepackage{color}
\usepackage{epstopdf}
\usepackage[version=3]{mhchem}
\usepackage{lipsum}
\usepackage[outercaption]{sidecap}
\usepackage{floatrow}
\usepackage{hyperref}
\usepackage{multirow}

\begin{document}

\title{Machine Learning-Integrated Modeling of Thermal Properties and Relaxation Dynamics in Metallic Glasses}

\author{Ngo T. Que}
\affiliation{Phenikaa Institute for Advanced Study, Phenikaa University, Hanoi 12116, Vietnam}
\author{Anh D. Phan}
\email{anh.phanduc@phenikaa-uni.edu.vn}
\affiliation{Faculty of Materials Science and Engineering, Phenikaa University, Hanoi 12116, Vietnam}
\affiliation{Phenikaa Institute for Advanced Study, Phenikaa University, Hanoi 12116, Vietnam}
\author{Truyen Tran}
\affiliation{Applied Artificial Intelligence Institute (A2I2), Deakin University, Australia}
\author{Pham T. Huy}
\affiliation{Faculty of Materials Science and Engineering, Phenikaa University, Hanoi 12116, Vietnam}
\author{Mai X. Trang}
\affiliation{Faculty of Computer Science, Phenikaa University, Hanoi 12116, Vietnam}
\author{Thien V. Luong}
\affiliation{Business AI Lab, Faculty of Data Science and Artificial Intelligence, National Economics University, Hanoi 10000, Vietnam}
\date{\today}

\begin{abstract}
Metallic glasses are a promising class of materials celebrated for their exceptional thermal and mechanical properties. However, accurately predicting and understanding the melting temperature ($T_m$) and glass transition temperature ($T_g$) remains a significant challenge. In this study, we present a comprehensive approach that integrates machine learning (ML) models with theoretical methods to predict and analyze these key thermal properties in metallic glasses. Our ML models using distributional data derived from elemental composition-based features obtain high accuracy while minimizing data preprocessing complexity. Furthermore, we explore the correlation between $T_m$ and $T_g$ to elucidate their dependence on alloy composition and thermodynamic behavior. When the $T_g$ value of metallic glasses is known, further analysis using the Elastically Collective Nonlinear Langevin Equation (ECNLE) theory provides a deeper understanding of structural relaxation dynamics. This integrated framework establishes a quantitative description consistent with experimental data and previous works and paves the way for the efficient design and discovery of advanced materials with tailored thermal properties.
\end{abstract}

\maketitle
\section{Introduction}
Metallic glasses (MGs) are amorphous alloys known for their exceptional properties, including high strength, excellent corrosion resistance, and unique thermal and magnetic behaviors \cite{ref1, ref2, ref3, ref4}. These characteristics have garnered significant attention in both scientific research and practical applications, spanning fields such as aerospace, biomedical devices, and electronics \cite{ref1, ref2, ref3, ref4}. The thermodynamic and kinetic stability of MGs is primarily governed by two critical temperatures: the melting temperature and the glass transition temperature. The melting temperature marks the transition from a solid to a liquid state, while the glass transition temperature defines the transformation from a glassy state to a supercooled liquid. These temperatures directly influence essential material properties, including thermal stability, mechanical performance, and glass-forming ability (GFA) \cite{ref3, 40,41}.

Accurate prediction of $T_m$ and $T_g$ is essential for the design and optimization of MGs \cite{10, 11, 12, GFA_1}. The $T_g$ and $T_m$ ratio ($T_g/T_m$) often referred to as the reduced glass transition temperature is a key parameter in determining GFA \cite{GFA_1, GFA_2, GFA_3_Tg}. Despite their importance, the experimental determination of $T_m$ and $T_g$ is often time-consuming and resource-intensive. Experimental techniques such as differential scanning calorimetry (DSC) and broadband dielectric spectroscopy (BDS) require extensive effort, particularly for exploring compositional spaces in multi-component alloys \cite{ref1, ref2, ref3, ref4}. Moreover, understanding the correlation between $T_m$ and $T_g$ remains a challenge due to their dependence on complex factors including atomic size mismatch, electronegativity differences, and cooling rates.

Recent advancements in machine learning (ML) and deep learning (DL) have provided powerful tools for efficiently predicting critical material properties. Machine learning models have successfully predicted various MG characteristics, including GFA, elastic moduli, and crystallization temperatures \cite{GFA_1, GFA_2, GFA_3_Tg, Bulk_1, Bulk_2, 2, 3, 4, 5, 7, 8}. Applying these methods to predict $T_m$ and $T_g$ introduces unique challenges, as these predictions often rely on complex features such as atomic radius \cite{3, 4}, valence electron concentration \cite{8}, and electrical resistivity \cite{8}. Simplifying these input features without compromising accuracy remains a critical research area. Additionally, dataset quality and diversity, often sourced from experimental data or atomistic simulations, significantly influence model performance and generalizability. While some studies have explored simpler input features \cite{dataset_Rao}, their focus has been limited to other properties, such as thermal expansion coefficients, rather than $T_m$ and $T_g$. 

This raises important research questions: (1) Can input features for ML models be simplified compared to the complex and extensive descriptors used in Refs. \cite{GFA_1, GFA_2, GFA_3_Tg, Bulk_1, Bulk_2, 2, 3, 4, 5, 7, 8} while maintaining predictive accuracy? (2) Is it feasible to predict the melting temperature using only alloy composition without the need for extensive physical properties or derived features? (3) Since most existing methods \cite{GFA_1, GFA_2, GFA_3_Tg, Bulk_1, Bulk_2, 2, 3, 4, 5, 7, 8} rely on a single type of dataset, but can the integration of experimental data, density functional theory (DFT) results, molecular dynamics (MD) simulations, and thermodynamic calculations improve $T_m$ prediction accuracy? Additionally, none of the current methods have successfully predicted the glass transition temperature of metallic glasses based solely on chemical composition, nor have they determined the temperature dependence of structural relaxation times from alloy composition alone. This leads to further questions: (4) Can ML models with simplified features determine $T_g$ effectively? (5) Can such approaches predict the structural relaxation time’s temperature dependence and provide quantitative compare with BDS data? Addressing these questions will advance the design and understanding of metallic glasses using streamlined and integrative ML approaches.

In this work, we address these challenges by developing predictive models for $T_m$ and $T_g$ using simplified input features to minimize preprocessing complexity. Our features are simply based on constituent elements of alloys. Then, we employ six widely used ML models to predict $T_m$ and $T_g$ across a diverse range of metallic glasses. Additionally, we analyze the correlation between $T_m$ and $T_g$ to elucidate their dependence on composition and thermodynamic behavior. After obtaining the $T_g$ value, we apply the ECNLE theory to determine the temperature-dependent structural relaxation times \cite{13,14,15,16,17,18,19,20,21,22,23}. This integrated framework provides a comprehensive and data-driven approach for predicting and analyzing the thermal properties of metallic glasses. This approach can be exploited to investigate high-entropy alloys and other amorphous alloys and design advanced materials with tailored thermal and mechanical properties.

\section{Method}
\begin{figure*}[htp]
\includegraphics[width=16cm]{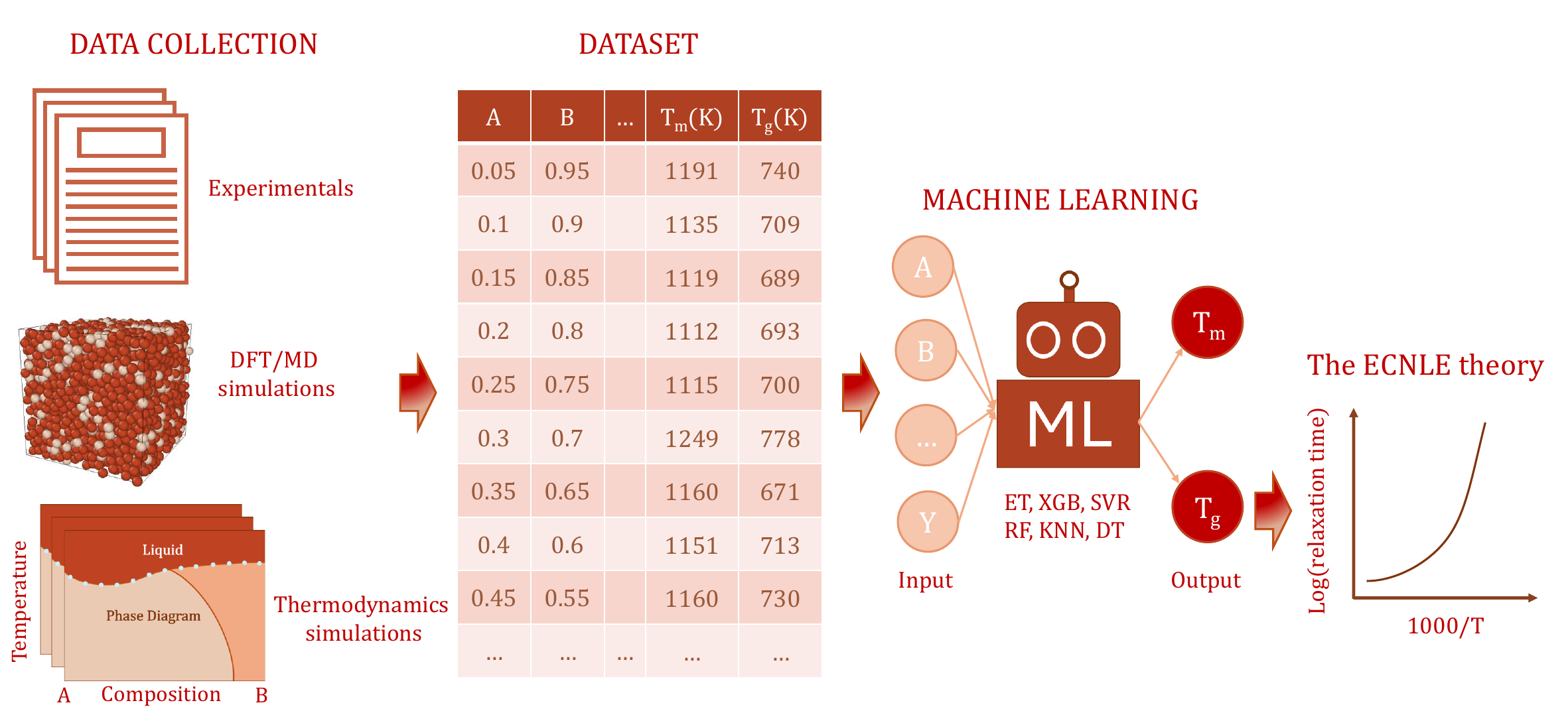}
\caption{\label{fig:0}(Color online) Workflow for predicting the melting temperature, the glass transition temperature, and subsequently the temperature dependence of the alpha relaxation time.}
\end{figure*}

The entire modeling process consists of four steps as illustrated in Fig. \ref{fig:0}. The first step involves collecting a suitable dataset. The second step is processing the dataset with input features to obtain the desired output, which includes modeling with the features used to predict \(T_m\) and \(T_g\). Finally, the ML-based $T_g$ value of MG can be exploited to calculate the temperature dependence of the relaxation time of metallic glasses in the final step.

\subsection{Data collection}
In this study, we utilized three datasets to develop and validate our predictive models. First, we employed an experimental dataset from Ref. \cite{GFA_1}, which contains data on 715 metallic glasses. This dataset includes chemical compositions and 21 parameters such as $T_m$, $T_g$, and various physical properties. The materials comprise 42 elements with Zr, Cu, Mg, Ni, and Cr as common constituents. Second, we used a dataset derived from thermodynamic calculations in Ref. \cite{dataset_Rao}. This dataset, generated using Thermo-Calc 2019b with the High Entropy Alloys v.2.1.1 database, employed CALPHAD (CALculation of PHAse Diagrams) calculations to determine phase equilibria and thermodynamic properties of face-centered cubic (FCC) Fe-Co-Ni-Cr-V-Cu alloys. After the calculations were performed across a range of compositions, the melting temperatures were extracted from the resulting phase diagrams. This dataset comprises 699 entries with each representing a material composition and its $T_m$. Finally, we constructed a new computational dataset generated by pyCALPHAD calculations \cite{9}. Our dataset contains 1240 data points for binary and ternary alloys which cover phase diagrams for 16 alloy systems \cite{9}. The alloys consist of two or three elements selected from Fe, Ni, Co, Cr, and Al with composition ratios varying in increments of 0.05. These simulations were performed within a temperature range of 500 K to 3000 K at 101325 Pa pressure and used Gibbs free energy minimization to precisely determine the melting temperatures of the alloys. All datasets generated and analyzed in this study are available in the Supplementary Material (Table S1, S2 and S3) \cite{50}.

\subsection{Features engineering}
After collecting these datasets, we processed and transformed them into inputs suitable for the predictive model. Each chemical formula was represented in a standardized form, \(A_aB_b\ldots Y_y\), where \(A\), \(B\), and \(Y\) denote elements, and \(a\), \(b\), and \(y\) correspond to their respective atomic percentages in the alloy to ensure that $(a + b + \ldots + y =$ 100 $\%$. This is called distributional data which allows the chemical composition of each alloy to be represented as an array of 42 fixed features, where the presence or absence of an element was indicated by its corresponding percentage. Elements absent in a given alloy were assigned a value of \(0\). 

Table~\ref{tab:addlabel} presents examples of the processed data structure used as input for our predictive models. Five metallic glass alloys are shown with each row representing a single alloy and containing its chemical composition expressed as elemental percentages, $T_g$, $T_m$. For example, the row corresponding to \ce{Cu_{60}Zr_{33}Ti_7} shows $T_g =$ 740 K, $T_m =$ 1191 K, and elemental percentages of 33 $\%$ for Zr, 7 \ce{\%} for Ti, 60 $\%$ for Cu, and 0 $\%$ for all other listed elements (La, Co, Ag, C, etc.). This format, where each element is represented by a dedicated column, facilitates direct input into the machine-learning algorithms.
 
\begin{table*}[htbp]
  \centering
  \caption{Examples of processed data structures and corresponding features for five metallic glass alloys including $T_g$, $T_m$, and chemical compositions.}
    \begin{tabular}{|l|c|c|c|c|c|c|c|c|c|c|c|}
    \hline
    Alloy & $T_g \ce{(K)}$        & $T_m  \ce{(K)}$    & Zr    & Ti    & Al    & Cu    & La    & …     & Co    & Ag    & C \\
    \hline
    \ce{Cu_{60}Zr_{33}Ti_7} & 740      & 1191  & 33    & 7     & 0     & 60    & 0     & …     & 0     & 0     & 0 \\ \hline
    \ce{Cu_{54}Ag_6Zr_{33}Ti_7} & 709      & 1135  & 33    & 7     & 0     & 54    & 0     & …     & 0     & 6     & 0 \\ \hline
    \ce{Cu_{46.4}Ag_{11.6}Zr_{35}Ti_7} & 689      & 1119  & 35    & 7     & 0     & 46.4  & 0     & …     & 0     & 11.6  & 0 \\ \hline
    \ce{Cu_{44.25}Ag_{14.75}Zr_{35}Ti_6} & 693      & 1112  & 35    & 6     & 0     & 44.25 & 0     & …     & 0     & 14.75 & 0 \\ \hline
    \ce{Cu_{44.25}Ag_{14.75}Zr_{36}Ti_5} & 700     & 1115  & 36    & 5     & 0     & 44.25 & 0     & …     & 0     & 14.75 & 0 \\
    \hline
    \end{tabular}%
  \label{tab:addlabel}%
\end{table*}%

\subsection{Machine learning modeling}
To comprehensively investigate the relationships between composition and thermo-properties in metallic glasses, we employed six well-established machine learning models: Extra Trees (ET), Gradient Boosting (XGB), Support Vector Regression (SVR), Random Forest (RF), K-Nearest Neighbors (KNN), and Decision Tree Regressor \cite{38}. These models were selected for their demonstrated high predictive performance in prior studies on metallic glasses \cite{42,44,45,46} and high-entropy alloys \cite{43}, as well as their capability to capture complex composition-property relationships. Tree-based ensemble methods (ET, XGB, RF) are particularly well-suited for capturing intricate relationships within high-dimensional data, while SVR offers advantages in handling non-linear data through kernel functions. KNN provides a complementary approach by considering local data patterns. Compared to deep learning models, these machine learning approaches are significantly simpler and require fewer parameters and less computational power while maintaining strong predictive capabilities \cite{47}. Moreover, they offer greater interpretability and are more suitable for smaller datasets, which is critical for materials science applications where experimental data is often limited \cite{38,42,43,44,45,46,47}. For example, feature importance scores in tree-based methods can explicitly quantify the influence of specific elements on the glass transition and melting temperatures. Similarly, SVR maps the data to a high-dimensional space and determines an optimal hyperplane for separating data points. The support vectors can define key compositional boundaries that influence properties, with their positions and weights indicating the impact of different alloying elements on thermo-properties. However, extracting physical meanings from these model outputs, such as SVR support vector weights or tree-based feature importance, requires extensive validation against experimental or simulation data to ensure reliability and prevent misinterpretation. Since the primary goal of this study is to establish an accurate predictive framework for $T_g$, $T_m$, and the temperature dependence of the structural relaxation time, we focus on predictive performance rather than an in-depth interpretability analysis. To maximize predictive performance, hyperparameter optimization was performed using random search. The dataset was randomly partitioned into training and testing subsets using an 80:20 split. The performance of the model was evaluated using the determination coefficient ($R^2$) and the mean squared error (RMSE) calculated as follows 
\begin{equation}
R^2 = 1 - \frac{\sum_{i=1}^{n}(y_i - \hat{y}_i)^2}{\sum_{i=1}^{n}(y_i - \bar{y})^2},
\end{equation}
\begin{equation}
\ce{RMSE} = \sqrt{\frac{1}{n} \sum_{i=1}^{n}(y_i - \hat{y}_i)^2},
\end{equation}
where $y_i$ and $\hat{y}_i$ are the observed and predicted value for the i-th data point, $\bar{y}$ is the mean of the observed value, and $n$ is the number of data points. Higher $R^2$ values, combined with lower RMSE and $\sigma$ values, indicate improved predictive accuracy.

\subsection{Predicting the temperature dependence of structural relaxation time}
After determining $T_g$, the ECNLE theory can be applied to investigate the temperature dependence of structural relaxation dynamics and related properties in the supercooled liquid regime \cite{13,14,15,16,17,18,19,20,21,22,23}. This theory describes glass-forming liquids as a hard-sphere fluid characterized by the particle diameter, $d$, density number, $\rho$, and the volume fraction, $\Phi = \rho \pi d^3/6$. The radial distribution function, $g(r)$, and static structure factor, $S(q)$ (where $q$ is the wavevector and $r$ is the radial distance), are determined using the Percus-Yevick (PY) integral equation theory. In metallic glasses, the total radial distribution function is derived as a composition-weighted sum of partial radial distribution functions from all constituent atomic pairs. To determine the total radial distribution function, we first calculate the partial radial distribution functions for each pair of elements, normalize them, and then combine them by weighting each one with the product of the atomic concentrations of the two elements involved. In the ECNLE framework, the particle diameter $d$ can be approximately defined by the location of the first peak in the radial distribution function which represents the first coordination shell and typically ranges from 2 to 4 $\AA$. This value corresponds to the average distance between neighboring atoms in the amorphous structure of metallic glasses \cite{51}.

The dynamics of a tagged particle is governed by interactions with its nearest neighbors and cooperative motion of the surrounding fluid. The local dynamics, describing the tagged particle's motion within its cage, is quantified by the dynamic free energy, $F_{\text{dyn}}(r)$,  expressed as the sum of two contributions $F_{\text{dyn}}(r) = F_{\text{ideal}}(r) + F_{\text{caging}}(r)$. Here, $F_{\text{ideal}}(r)=-3k_BT\ln\left(r/d\right)$ represents the free energy of the delocalized (ideal fluid) state, while $F_{\text{caging}}(r)$ captures the localized (caged) state reflecting the influence of density and structural order. Here, $k_B$ is Boltzmann constant and $T$ is the temperature. Detailed calculations of $F_{\text{caging}}(r)$, utilizing the volume fraction and $S(q)$, are presented in previous works \cite{13,14,15,16,17,18,19,20,21,22,23}.

\begin{figure}[htp]
\includegraphics[width=8cm]{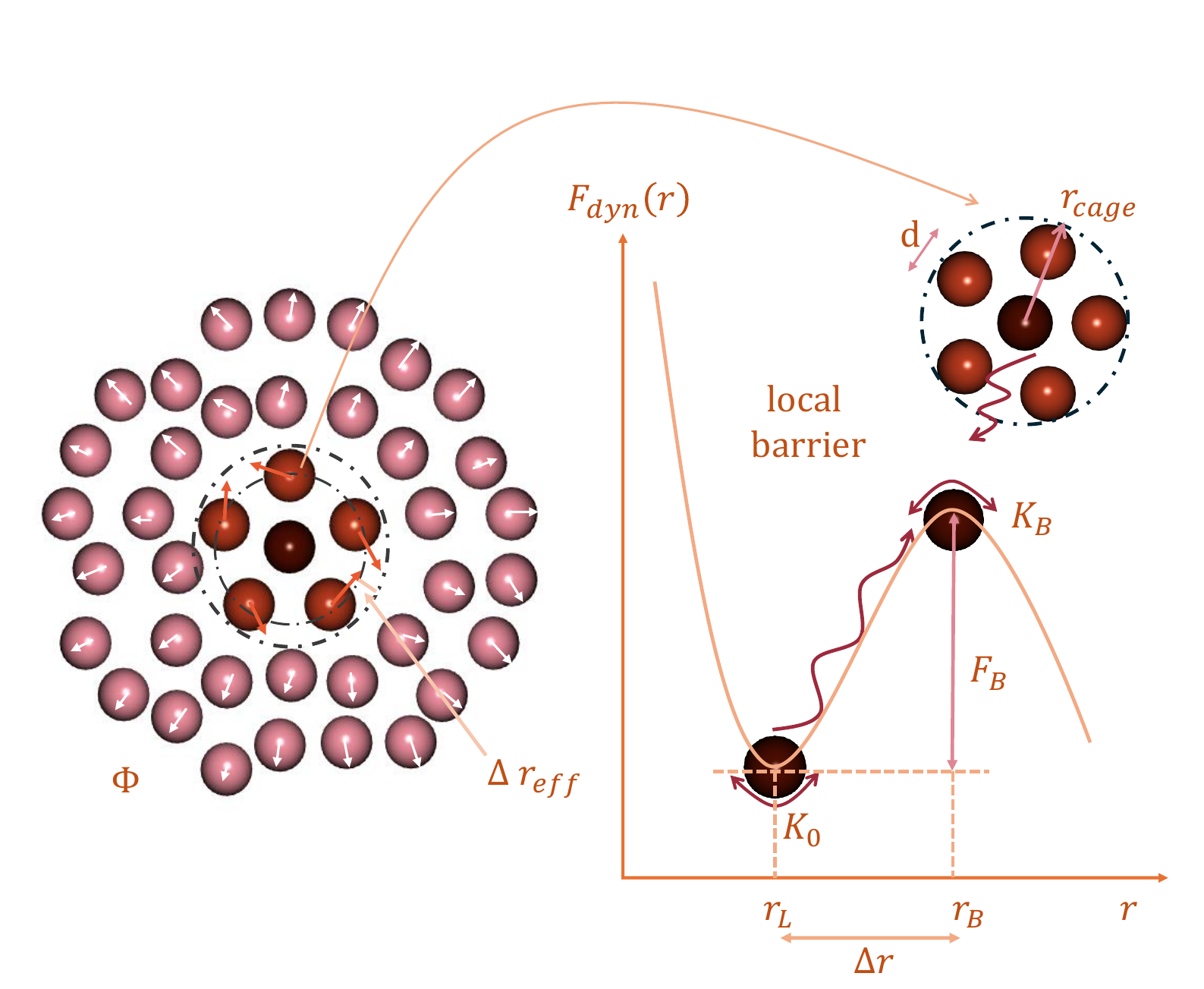}
\caption{\label{fig:ECNLE}(Color online) Schematic illustration of the dynamic free energy in the ECNLE theory. Key length and energy quantities are defined.}
\end{figure}

At sufficiently high densities, the free volume is reduced and the tagged particle becomes dynamically confined within a cage formed by its neighboring particles. The onset of transient localization occurs as a barrier emerges in $F_{\text{dyn}}(r)$. Then, we can calculate the key length and energy scales of the local dynamics as depicted in Fig. \ref{fig:ECNLE}. The local minimum of $F_{dyn}(r)$ defines the localization length, $r_{\text{L}}$, while the barrier position is denoted by  $r_{\text{B}}$. From these, the jump distance and the local barrier height are determined by $\Delta r = r_{\text{B}} - r_{\text{L}}$ and $F_{\text{B}} = F_{\text{dyn}}(r_{\text{B}}) - F_{\text{dyn}}(r_{\text{L}})$, respectively. In addition, we calculate the harmonic curvatures at $r_{\text{L}}$ and $r_{\text{B}}$ as $K_0 = \left.\cfrac{\partial^2 F_{\text{dyn}}(r)}{\partial r^2}\right|_{r = r_{\text{L}}}$ and $K_{\text{B}} = \left|\cfrac{\partial^2 F_{\text{dyn}}(r)}{\partial r^2}\right|_{r = r_{\text{B}}}$, respectively. $K_0$ represents the effective spring constant at the localization length. 

For many amorphous materials \cite{13,14,15,16,17,18,19,20,21}, cage escape necessitates cooperative rearrangements of neighboring and surrounding particles with these collective motions coupled to local dynamics. However, in metallic glasses, effects of collective motion on the glass transition are considered negligible with only local dynamics playing the dominant role \cite{22,23}. Consequently, incorporating only the local barrier into Kramer's theory provides a reasonable approximation for the structural relaxation time \cite{13,14,15,16,17,18,19,20,21,22,23}
\begin{equation}
    \frac{\tau_\alpha}{\tau_s} = 1 + \frac{2\pi}{\sqrt{K_0 K_{\text{B}}}} \frac{k_{\text{B}} T}{d^2} \exp\left( \frac{F_{\text{B}}}{k_{\text{B}} T} \right),
    \label{eq:4}
\end{equation}
where $\tau_s$ is a short-time scale defined elsewhere \cite{13,14,15,16,17,18,19,20,21,22,23}. Equation (\ref{eq:4}) provides the density dependence of the structural relaxation time, $\tau_\alpha(\Phi)$. For quantitative comparison with experimental data, we use the density-to-temperature conversion constructed using the thermal expansion process \cite{13,14,15,16,17,18,19,20,21,22,23}
\begin{equation}
    T = T_g + \frac{\Phi_g - \Phi}{\beta \Phi_0},
    \label{eq:5}
\end{equation}
where $T_g$ is the glass transition temperature defined as $\tau_\alpha(T_g) = 100$ s given by simulations, experiment, or ML prediction. $\Phi_g \approx 0.6714$ is the volume fraction at $\tau_\alpha(\Phi_g) = 100$ s for metallic glasses, $\Phi_0 \approx 0.50$ is a characteristic volume fraction, and $\beta \approx 12 \times 10^{-4} \, \text{K}^{-1}$ is an effective thermal expansion coefficient \cite{13,14,15,16,17,18,19,20,21,22,23}.
\section{Results AND Discussion}
\subsection{Predicting the melting temperature}
\begin{figure*}[htp]
\includegraphics[width=16cm]{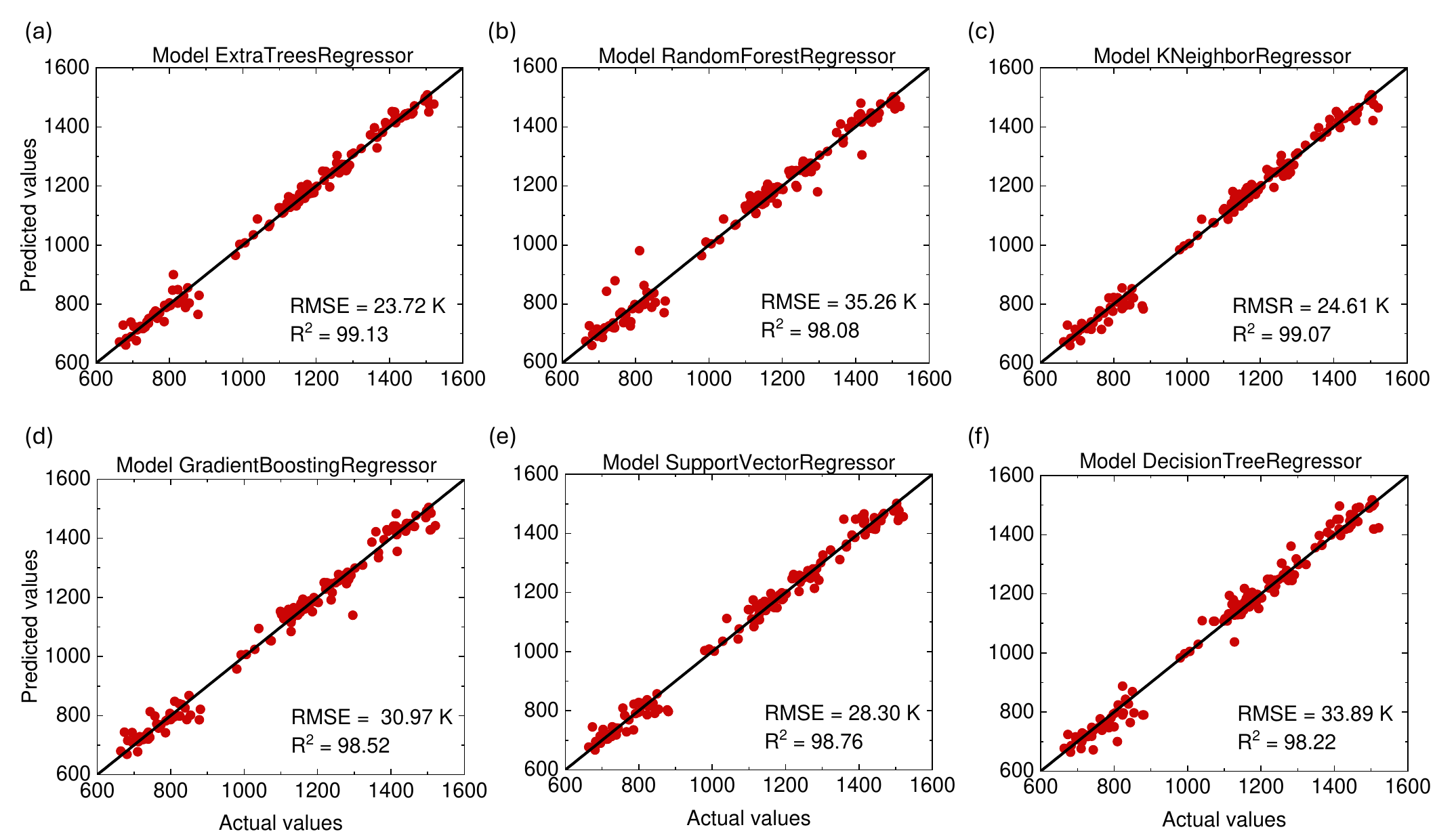}
\caption{\label{fig:1}(Color online) Predictive performance of six regression models including (a) Extra Trees, (b) Random Forest, (c) Gradient Boosting, (d) Support Vector Regression, (e) K-Nearest Neighbors, and (f) Decision Tree on the testing dataset for the $T_m$ prediction using Ghorbani's experimental dataset \cite{GFA_1}. The models were evaluated based on RMSE and $R^2$.}
\end{figure*}

Figure~\ref{fig:1} shows the performance of various machine learning models in predicting \(T_m\) for the experimental metallic glass dataset \cite{GFA_1}. Among these, the Extra Trees Regressor demonstrates the highest predictive accuracy with an RMSE of 23.72 K and an \(R^2\) value of 99.13 \ce{\%} on the test set. Other models including Random Forest, Gradient Boosting, Support Vector Machine, and Decision Tree also exhibit strong predictive capabilities. The \(R^2\) values for these models range from 98.08 \ce{\%} to 99.07 \ce{\%}, while their RMSE values vary between 24.61 K and 35.26 K. This finding indicates that the Extra Trees Regressor captures better nonlinear interactions among alloy compositions. Meanwhile, the consistent results across other models indicate that the dataset's quality significantly reduces prediction variability and enhances reliability.

\begin{figure*}[htp]
\includegraphics[width=16cm]{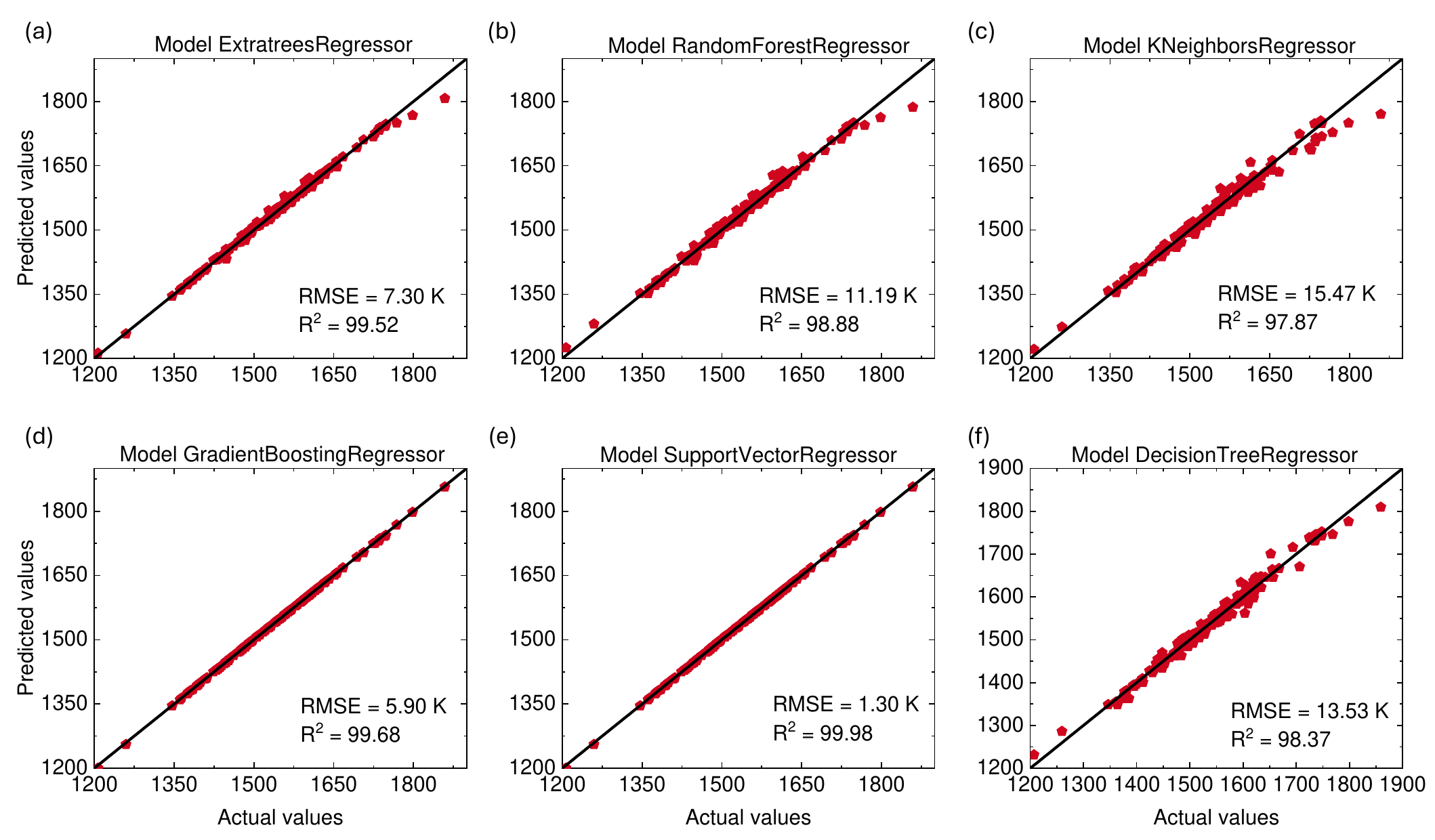}
\caption{\label{fig:2}(Color online) Predictive performance of six regression models including (a) Extra Trees, (b) Random Forest, (c) Gradient Boosting, (d) Support Vector Regression, (e) K-Nearest Neighbors, and (f) Decision Tree on the testing dataset for predicting $T_m$ of Fe-Co-Ni-Cr-V-Cu alloys using the dataset derived from CALPHAD calculations \cite{dataset_Rao}.}
\end{figure*}

Figure~\ref{fig:2} presents predictive results of the above six machine learning models for $T_m$ using a dataset derived from CALPHAD simulations \cite{dataset_Rao} of Fe-Co-Ni-Cr-V-Cu alloys. CALPHAD calculations predict stable phase equilibria by minimizing the system’s Gibbs free energy and ensuring that the generated compositions are thermodynamically stable at specific temperatures. While thermodynamic stability does not guarantee ease of synthesis under all experimental conditions, it provides strong evidence of their plausibility and accessibility for experimental realization. The results in Fig. \ref{fig:2} show high predictive accuracy across all models with \(R^2\) values ranging from 97.87 \ce{\%} to 99.98 \ce{\%} and RMSE values between 1.30 K and 15.47 K. The Support Vector Regressor achieves the best performance with the lowest RMSE of 1.30 K and an $R^2$ value of 99.98\%. The Gradient Boosting Regressor and Extra Trees Regressor also perform well with RMSE values below 7.30 K. The observed high performance across different models and datasets indicates effectiveness of machine learning for predicting material properties from both experimental and computational data. This finding is particularly important for materials discovery, where datasets often vary in origin and complexity.

\begin{figure*}[htp]
\includegraphics[width=16cm]{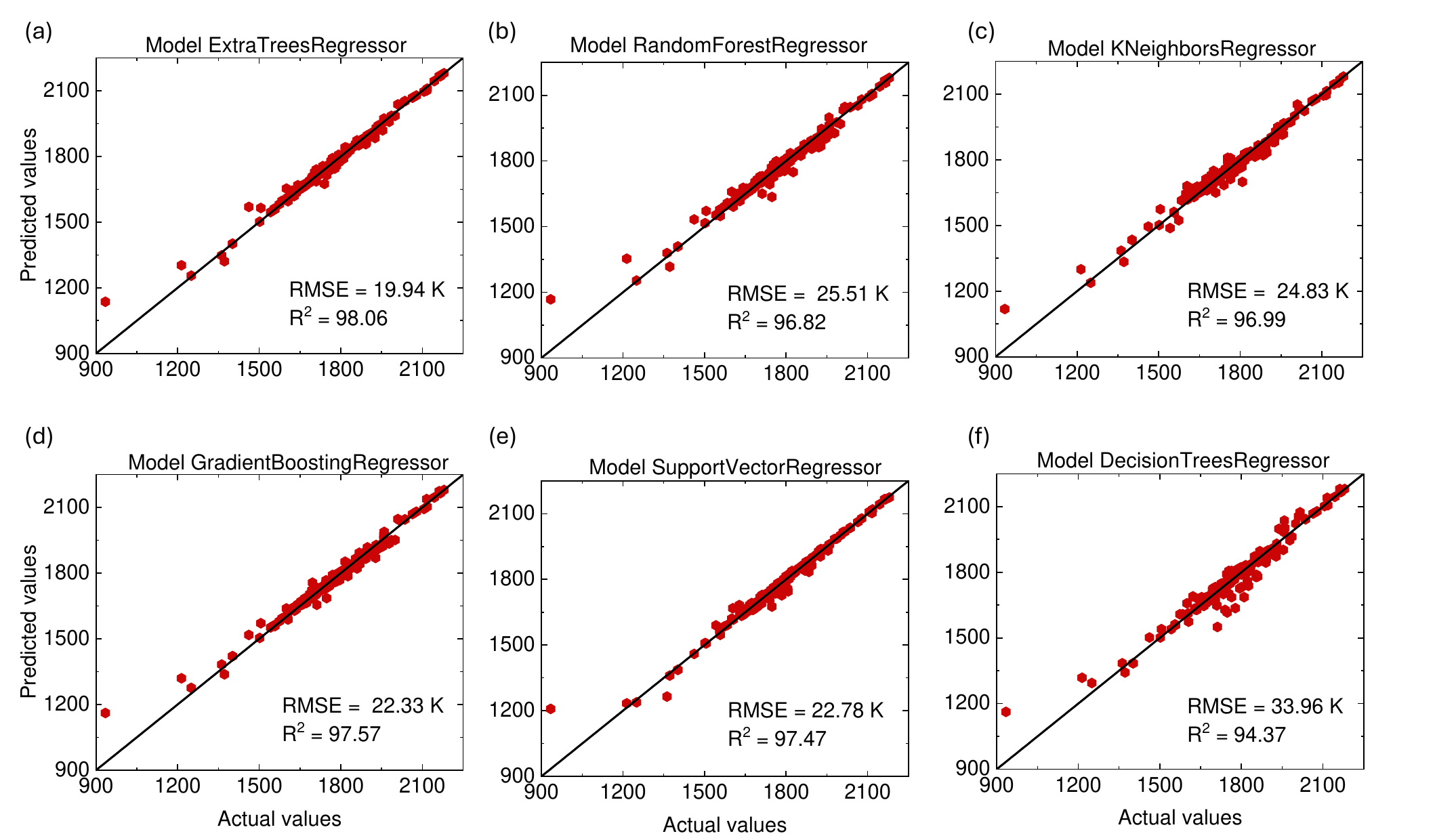}
\caption{\label{fig:3}(Color online) Predictive performance of six regression models including (a) Extra Trees, (b) Random Forest, (c) Gradient Boosting, (d) Support Vector Regression, (e) K-Nearest Neighbors, and (f) Decision Tree on the testing dataset for predicting $T_m$ of our 1240 binary and ternary alloys composed of Fe, Ni, Co, Cr, and Al.}
\end{figure*}

While the CALPHAD dataset from Ref.\cite{dataset_Rao} provides a valuable benchmark for evaluating model performance on complex multi-component systems, its composition space is relatively constrained. To address this limitation and investigate simpler alloy systems, we developed a new computational dataset using the open-source CALPHAD tool \cite{9}. This dataset includes binary and ternary alloys composed of Fe, Ni, Co, Cr, and Al with systematic variations in composition to be consistent with the Fe-Co-Ni-Cr-V-Cu system in Ref. \cite{dataset_Rao}. Figure \ref{fig:3} shows the predictive performance of the six machine learning models applied to this thermodynamic simulation dataset. The ML-based predictions quantitatively agree with the simulated melting temperatures with $R^2$ values ranging from 94.37 $\%$ to 98.06 $\%$. This agreement reflects a high correlation between predicted and simulated melting temperatures. Furthermore, RMSE values between 19.94 K and 33.96 K indicate minimal deviation. 

Compared to prior works, our models consistently outperform previous approaches in predictive accuracy. Table~\ref{tab:model_comparison} provides a comprehensive comparison of machine learning approaches applied to data from experiments, MD simulations, DFT simulations, and thermodynamic calculations. Hong and his co-workers \cite{2,7,37} employed graph neural networks (GNNs) and XGBoost to predict melting temperatures using extensive DFT-based datasets. Their models achieved $R^2$ scores of 91.9 \ce{\%}–93.3 \ce{\%} with RMSE values ranging from 138 K to 183 K. While their approach addresses a broader chemical space, our focus on alloys allows for greater precision in its predictions. Even when compared to studies using similar dataset sizes and machine learning models in Ref. \cite{5} which employs the Extra Trees model on 751 experimental data points and achieves an RMSE of 38.49 K and an $R^2$ of 98 \ce{\%}, our models working on 715 experimental and 699 thermodynamic data points achieve better accuracy with RMSE lower than 39 $K$ and $R^2$ greater than 97 $\%$. A key difference lies in the feature representation. While Ref. \cite{5} utilizes pre-calculated physicochemical properties as material descriptors, our approach uses elemental composition directly. These pre-calculated descriptors, being derived from other sources, inherently carry their own uncertainties. The application of Gradient Boosting to experimental datasets in Ref. \cite{3} resulted in a significantly larger RMSE of 194 K. The large deviation indicates the challenges of noisy real-world data originating from diverse measurements, techniques, experimental conditions, and samples. Although the MeltNet model in Ref. \cite{8}, trained on a very large thermodynamic dataset of 28148 entries, has an exceptionally low RMSE of 10.95 K, our models remain competitive, especially considering our substantially smaller dataset sizes (715 to 1240 entries). This comparison clearly shows the importance of data quality and effective feature engineering alongside dataset size in developing accurate predictive models.

\begin{table*}[h!]
\centering
\begin{tabular}{|c|l|l|c|c|l|}
\hline
\textbf{Size of Data} & \textbf{Type of Data} & \textbf{DL/ML Models} & \textbf{$R^2$} \ce{\%} & \textbf{RMSE} & \textbf{Reference} \\ \hline
9375 & DFT simulation & GNN & 93.3 & 160 K & \cite{2} \\ 
 &  & XGBoost & 91.9 & 183 K & \cite{2} \\
 &  & GNN + ResNet & - & 138 K & \cite{7} \\ \hline
476 & Experiment & Gradient Boosting & 92.0 & 194 K & \cite{3} \\ \hline 
593 & MD simulation & Random Forest & - & $>$ 36.44 K & \cite{4} \\ \hline
751 & Experiment & Extra Trees & 98.0 & 38.49 K & \cite{5} \\ \hline
28148 & Thermodynamics & MeltNet & - & 10.95 K & \cite{8} \\ \hline
715 & Experiment & ML models & $> 98.0$ & $<39$ K & This Work \\ \hline
699 & Thermodynamics &  ML models & $>97.8$ & $<15.5$ K & This Work \\ \hline
1240 & Thermodynamics & ML models & $>97.3$ & $<34$ K & This Work \\ \hline
\end{tabular}
\caption{Comparison of datasets, machine learning models, and performance metrics for the melting temperature prediction in this study and previous works \cite{2, 3, 4, 5, 7, 8}.}
\label{tab:model_comparison}
\end{table*}

\begin{figure*}[htp]
\includegraphics[width=16cm]{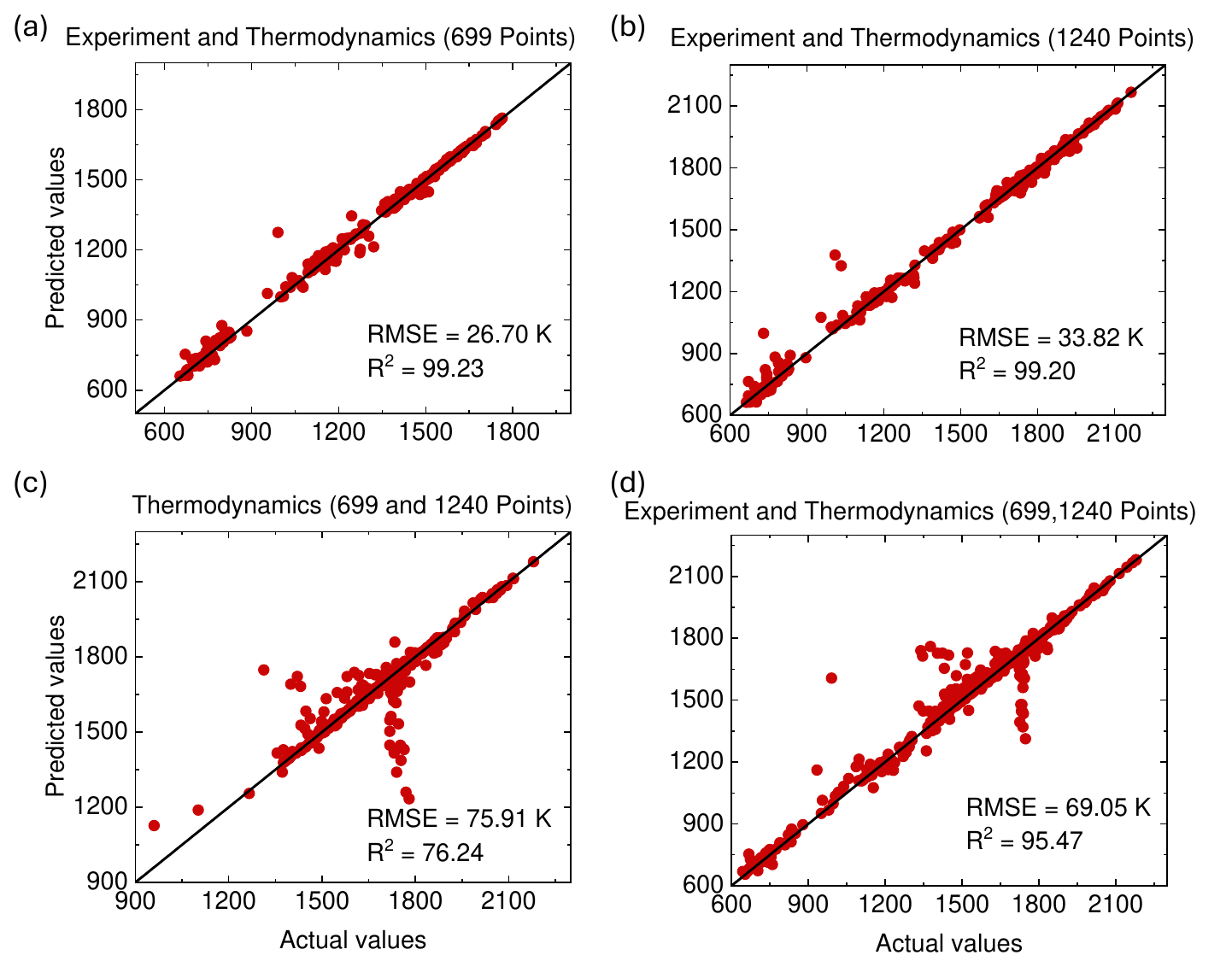}
\caption{\label{fig:4}(Color online) Evaluation of testing set predictions for melting temperatures using the Extra Trees Regressor with (a) Ghorbani's experimental dataset \cite{GFA_1} and Rao's CALPHAD calculations \cite{dataset_Rao}, (b) Ghorbani's experimental dataset \cite{GFA_1} and our thermodynamic data, (c) Rao's CALPHAD dataset \cite{dataset_Rao} and our thermodynamic data, (d) Ghorbani's and Rao's dataset \cite{GFA_1,dataset_Rao}, and our thermodynamic data plotted against actual values.}
\end{figure*}

From the above calculations, we find that the ExtraTrees Regressor effectively predicts the melting temperatures of multicomponent alloys across various datasets and temperature ranges. This raises a key question: can combining different datasets to create a larger and more comprehensive dataset further enhance predictive accuracy? Consequently, we merge the datasets used in Figs. \ref{fig:1}, \ref{fig:2}, and \ref{fig:3}, and retrain the ExtraTrees Regressor to determine $T_m$. Numerical results, as shown in Fig. \ref{fig:4}, reveal that combining Ghorbani's experimental data with either Rao's dataset or the CALPHAD-generated dataset developed in this study slightly improves the predictive accuracy with $R^2 > 99$ \ce{\%} and RMSE $\sim 27$ K. This suggests that data augmentation enhances both the accuracy and the analytical scope of the data. However, when our dataset is combined with Rao's data, the predictive accuracy declined significantly, despite both datasets being derived from CALPHAD calculations. This discrepancy likely stems from inconsistencies within the CALPHAD calculations themselves. Rao's dataset reports melting temperatures for pure iron and iron-based alloys around 1200 K, which is significantly lower than the established value of approximately 1811 K for pure iron and the expected range for most iron-based alloys. This suggests that an error in Rao's CALPHAD calculations potentially arises from the use of an incorrect or outdated thermodynamic database, an inappropriate thermodynamic model, or inaccurate model parameterization. Such discrepancies between the underlying thermodynamic descriptions negatively impact the performance of machine learning models trained on the combined dataset.

\subsection{Predicting the glass transition temperature and unveiling its link to the melting temperature}
\begin{figure*}[htp]
\includegraphics[width=16cm]{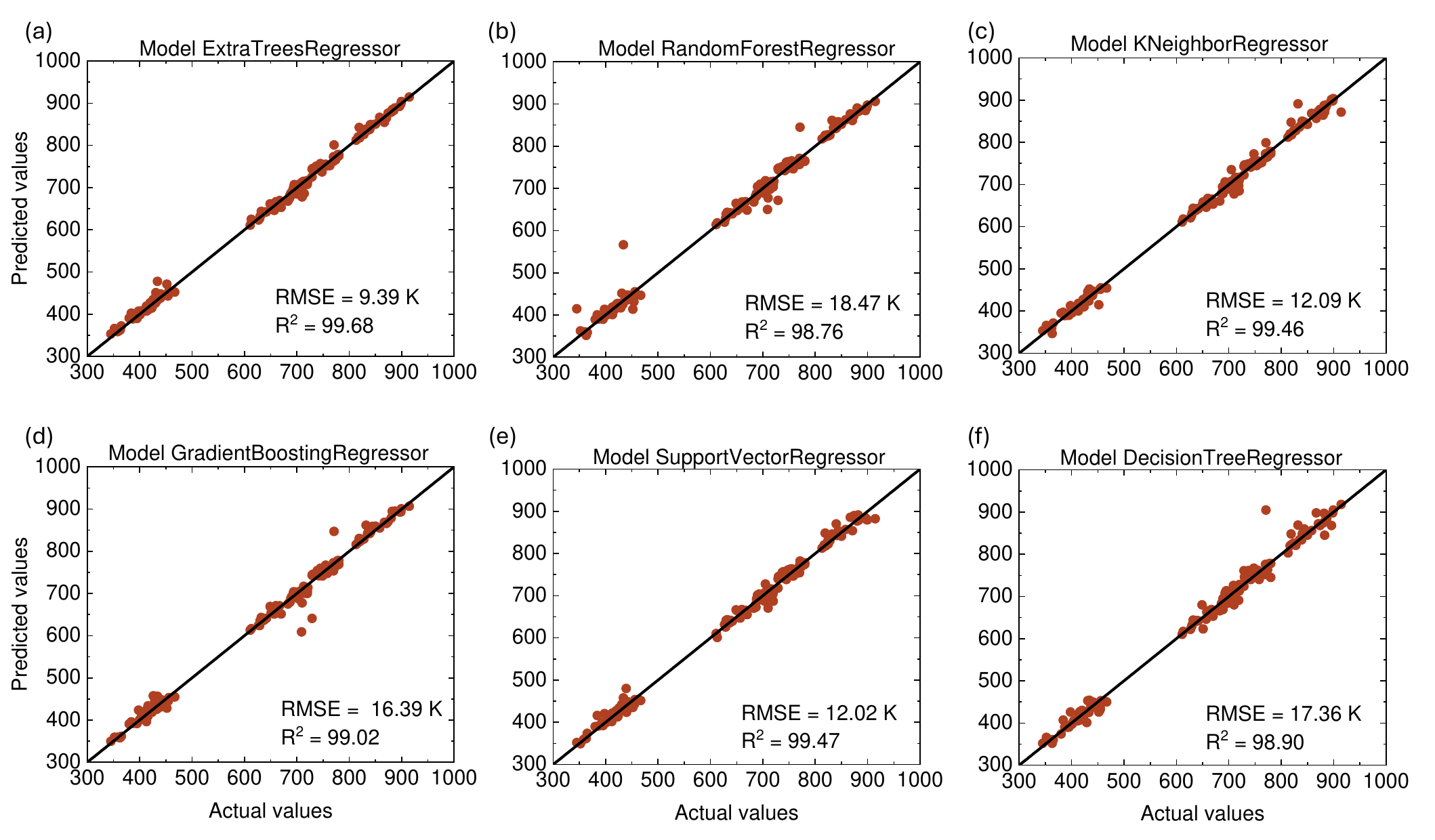}
\caption{\label{fig:5}(Color online) Predictive performance of six regression models including (a) Extra Trees, (b) Random Forest, (c) Gradient Boosting, (d) Support Vector Regression, (e) K-Nearest Neighbors, and (f) Decision Tree on the independent testing dataset for the $T_g$ prediction using Ghorbani's experimental dataset \cite{GFA_1}.}
\end{figure*}

Having established accurate predictions for melting temperatures, we extend the data-driven approaches to investigate their applicability in predicting glass transition temperatures. Using Ghorbani's experimental dataset \cite{GFA_1}, we retrain the six previously described machine learning models to determine $T_g$. Due to the limited availability of consistent experimental data on the $T_g$ dependence of metallic glasses on varying cooling rates, our machine-learning calculations focus on chemical composition and data obtained under standardized cooling rates (10-20 K/min). This approach allows for broad applicability and rapid material screening, and provides valuable initial estimations. Despite inherent complexities of $T_g$ prediction influenced by both compositional and structural factors, the models show strong predictive capabilities in Fig. \ref{fig:5}. The $R^2$ values exceed 98.76 \ce{\%} and the RMSE values are below 18.47 $K$. This indicates a stronger correlation between predicted and actual values for $T_g$ compared to the predictions for $T_m$. This difference in predictive accuracy may be attributed to the distinct nature of the two transitions. The melting process involves multiple crystalline phases and complex phase transitions. In contrast, the glass transition occurs in a single amorphous state which may simplify underlying physical relationships and reduce the complexity of predictions.

\begin{figure*}[htp]
\includegraphics[width=16cm]{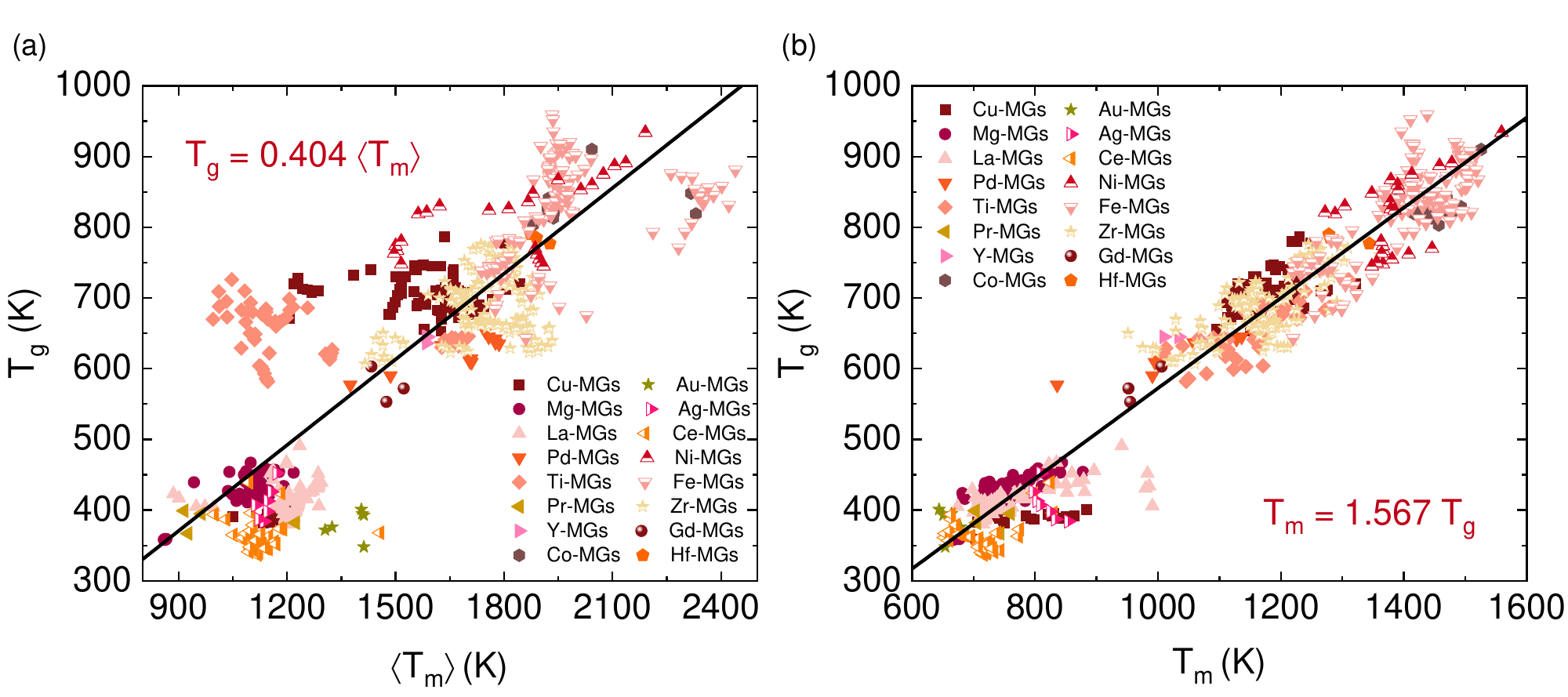}
\caption{\label{fig:6}(Color online) Correlation between (a) the average melting temperature, \( \langle T_m \rangle \), and (b) the melting temperature, \(T_m\), with $T_g$.}
\end{figure*}

A correlation between these properties is highly valuable for materials selection in various applications. In a prior work \cite{correlation_TgTm}, Lu and his coworkers analyzed a set of 210 bulk metallic glasses composed of 32 elements and established a linear relationship of \( T_g \approx 0.385 \langle T_m \rangle \) with \(\langle T_m \rangle\) being the average of melting temperature. An expression of \(\langle T_m \rangle\) is given by
\begin{equation}
    \langle T_m \rangle = \sum_i f_i T_{mi},
\end{equation}
where $f_i$ and $T_{mi}$ are the atomic fraction and melting temperature of the i-th component, respectively \cite{correlation_TgTm, meltingpoint}. We apply this approach to Ghorbani's dataset of 715 metallic glasses, which encompasses 42 elements, and compare the calculated $\langle T_m \rangle$ values with their corresponding $T_g$ values in Fig. \ref{fig:6}a. Our analysis presents a similar linear approximation $T_g \approx 0.404 \langle T_m \rangle$. However, significant deviations from this linear trend were observed with many materials falling outside the fitted line as shown in Fig. \ref{fig:6}a. These deviations indicate that the relationship between $T_m$ and $T_g$ is not merely governed by a constant scaling factor and suggests that other factors beyond average melting temperature influence the glass transition.

The ratio of $T_g/T_m$ known as the reduced glass transition temperature is a key indicator of a metallic glass's GFA. A higher $T_g/T_m$ corresponds to a reduced tendency for crystallization during cooling and thereby facilitates bulk glass formation. Understanding the factors influencing this ratio is therefore essential for the design of new bulk metallic glasses. To investigate this relationship within Ghorbani's dataset, we plot $T_g$ versus $T_m$ with the results shown in Fig. \ref{fig:6}b. Our analysis suggests a linear relationship of $T_m \approx 1.567 T_g$ for metallic glasses. This contrasts with our previous finding \cite{14} of $T_m \approx 1.362 T_g$ for polymers and amorphous drugs. This difference in the $T_m/T_g$ ratio between metallic glasses and polymeric/pharmaceutical amorphous materials possibly reflects fundamental differences in their bonding characteristics and atomic structures. Due to their dense atomic packing and strong interatomic bonding, metallic glasses require more energy to melt and this leads to a higher $T_m/T_g$ ratio than polymers and amorphous drugs, which have weaker intermolecular forces and more flexible chain dynamics.

\subsection{Predicting temperature-dependent structural relaxation using the ECNLE theory}

\begin{figure}[htp]
\includegraphics[width=9cm]{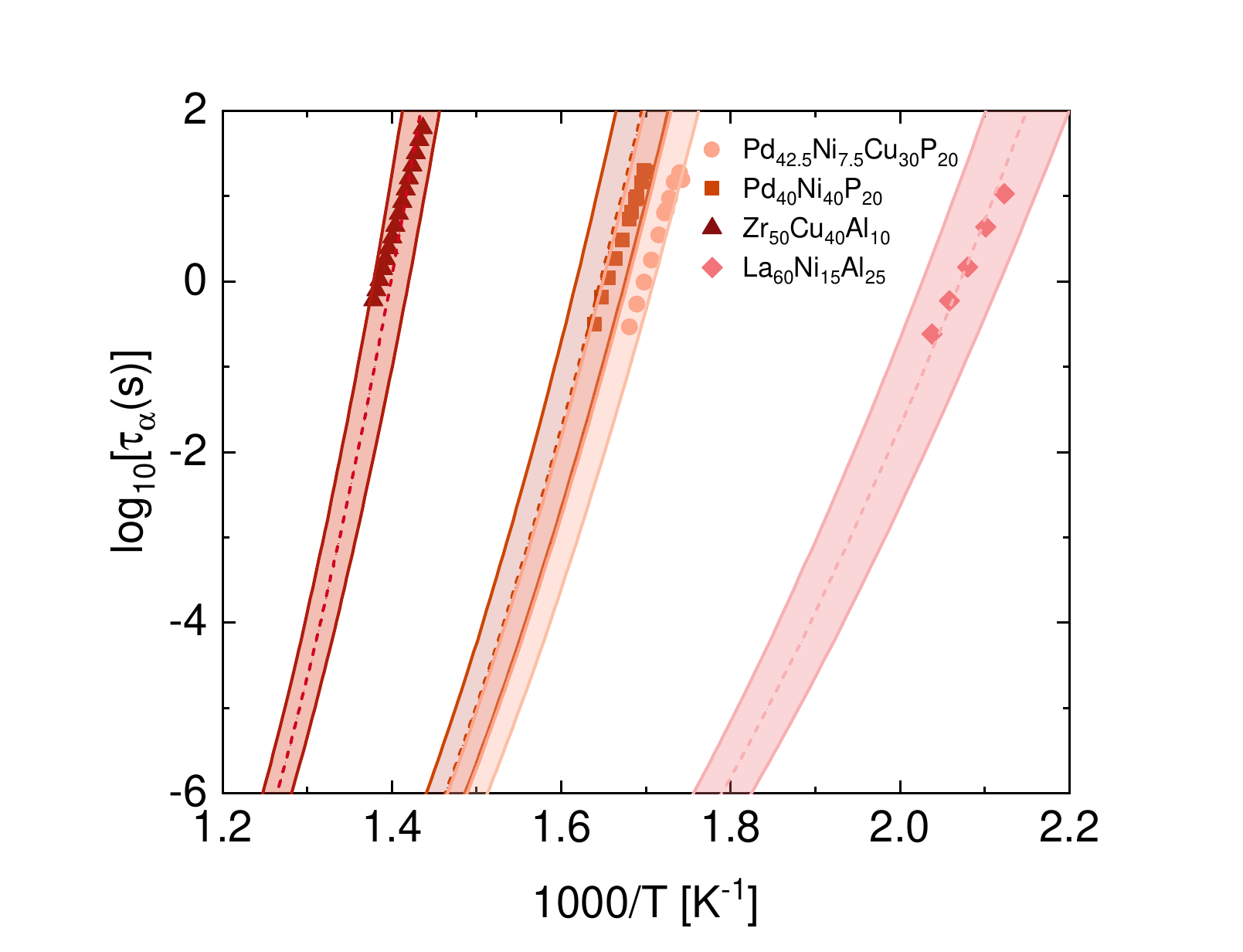}
\caption{\label{fig:7}(Color online) Logarithm of the experimental and theoretical structural relaxation time of several metallic glasses as a function of $1000/T$. Data points are experimental data in Refs.\cite{tau_PdNiP,tau_Zn38Mg12Ca32Yb18_Zr50Cu40Al10,tau_La60Ni15Al25} and dashed curves correspond to our ECNLE predictions for the temperature-dependent relaxation times using the $T_g$ values obtained from the ExtraTrees Regression model as determined from Fig. \ref{fig:5}. The shaded regions represent the range of relaxation times attributed to predictive model uncertainties, quantified by an RMSE of 10.66 K.}
\end{figure}

Given a predicted $T_g$ value for a metallic glass, the temperature-dependent structural relaxation time can be determined through ECNLE calculations using Eqs. (\ref{eq:4}) and (\ref{eq:5}). To the best of our knowledge, this is the first study to demonstrate the ability to predict the full temperature dependence of the structural relaxation time $\tau_\alpha(T)$ of metallic glasses directly from chemical composition. To validate the quantitative agreement between ECNLE predictions and experimental data, we performed calculations for four representative metallic glasses including \ce{Pd_{40}Ni_{40}P_{20}}, \ce{Pd_{42.5}Ni_{7.5}Cu_{30}P_{20}}, \ce{Zr_{50}Cu_{40}Al_{10}}, and \ce{La_{60}Ni_{15}Al_{25}}. As shown in Fig. \ref{fig:7}, while the predicted dynamic behavior shows a slight deviation in curvature compared to the experimental data, the structural relaxation times calculated using the ECNLE theory exhibit strong overall agreement with the experimental results across a wide range of metallic glass compositions. Remarkably, Notably, all experimental data points fall within the predicted confidence interval. This agreement indicates the capability of our integrated framework combining machine learning and the ECNLE theory to accurately predict $\tau_\alpha(T)$ based solely on material composition.  

\section{Conclusion}
We have developed the integrated computational framework combining ML models with ECNLE theory for predicting and analyzing key thermal properties of metallic glasses. By using only elemental composition as input features, our ML models achieve predictive accuracies of $R^2 > 99$ $\%$ for $T_m$ and $R^2 > 98.79$ $\%$ for $T_g$. These metrics outperform previous approaches \cite{2, 3, 4, 5, 7, 8} that relied on complex and extensive descriptors. This affirms that predictive models can effectively use minimal compositional data while maintaining accuracy. While Extra Trees methods exhibit superior accuracy, simpler models such as Decision Tree and KNN may be preferable when interpretability or lower computational cost is prioritized \cite{38,48,49}. To address the challenge of dataset consistency, we evaluated the impact of integrating different data sources. Combining Ghorbani's experimental data \cite{GFA_1} with our CALPHAD dataset improves $T_m$ prediction accuracy while merging Rao's dataset \cite{dataset_Rao} with ours decreases performance. This result clearly emphasizes the need for careful dataset integration in data-driven modeling. Additionally, we explored the correlation between $T_m$ and $T_g$ and established an improved linear relationship $T_m \approx 1.567T_g$. This result refines the traditional ratio of 1.5 with statistically significant precision.

Our approach also answers whether ML models can be exploited to predict temperature-dependent structural relaxation times directly from composition. By applying the ECNLE theory with ML-predicted $T_g$ values, we have, for the first time, accurately predicted the temperature-dependent structural relaxation times across various metallic glass systems. Our numerical calculations exhibit strong quantitative agreement with experimental data \cite{tau_PdNiP, tau_Zn38Mg12Ca32Yb18_Zr50Cu40Al10, tau_La60Ni15Al25}. In summary, this comprehensive framework addresses the critical research questions raised by demonstrating that minimal-feature ML models can reliably predict both $T_m$ and $T_g$ and link composition to relaxation dynamics. This methodology advances data-driven materials design and provides a scalable pathway for optimizing metallic glasses, high-entropy alloys, and other amorphous systems with tailored thermal properties.

\begin{acknowledgments}
The work of Anh D. Phan was supported by the Vietnam National Foundation for Science and Technology Development(NAFOSTED) under Grant No. 103.01-2023.62
\end{acknowledgments}
\section*{Conflicts of interest}
The authors have no conflicts to disclose.

\section*{Data availability}

The data that support the findings of this study are available from the corresponding author upon reasonable request.

\end{document}